\documentclass[twocolumn]{aastex62}

\usepackage[autostyle]{csquotes}

\shorttitle{Heliospheric modulation of Positrons}
\shortauthors{Aslam et al.}

\begin{document}

\title{Modeling of Heliospheric Modulation of Cosmic-ray Positrons in a Very Quiet Heliosphere}

\email{aslamklr2003@gmail.com (OPMA), driaanb@gmail.com (DB), \\
marius.potgieter@nwu.ac.za (MSP)}

\author[0000-0001-9521-3874]{O.P.M. Aslam}
\affil{Centre for Space Research, North-West University, 2520 Potchefstroom, South Africa \\}

\author[0000-0001-7623-9489]{D. Bisschoff}
\affil{Centre for Space Research, North-West University, 2520 Potchefstroom, South Africa \\}

\author[0000-0003-0793-7333]{M. S. Potgieter}
\affil{Centre for Space Research, North-West University, 2520 Potchefstroom, South Africa \\}

\author[0000-0002-8015-2981]{M. Boezio}
\affil{INFN, Sezione di Trieste I-34149 Trieste, Italy \\}

\author[0000-0001-7598-1825]{R. Munini}
\affil{INFN, Sezione di Trieste I-34149 Trieste, Italy \\}

\begin{abstract}

Heliospheric modulation conditions were unusually quite during the last solar minimum activity between Solar Cycles 23/24. Fortunately, the $\it{PAMELA}$ space-experiment measured six-month averaged Galactic positron spectra for the period July 2006 to December 2009, over an energy range of 80 MeV to 30 GeV, which is important for solar modulation. The highest level of Galactic positrons was observed at Earth during the July-December 2009 period. A well-established, comprehensive three-dimensional (3D) numerical model is applied to study the modulation of the observed positron spectra. This model had been used previously to understand the modulation of Galactic protons and electrons also measured by $\it{PAMELA}$ for the same period. First, a new very local interstellar spectrum for positrons is constructed, using the well-known GALPROP code together with the mentioned $\it{PAMELA}$ observations. The 3D model is used to distinguish between the dominant mechanisms responsible for the heliospheric modulation of Galactic positrons, and to understand the effect of particle drift during this unusual minimum in particular, which is considered diffusion dominant, even though particle drift still had a significant role in modulating positrons. Lastly, the expected intensity of Galactic positrons during an A$>$0 polarity minimum, with similar heliospheric conditions than for 2006-2009, is predicted to be higher than what was observed by $\it{PAMELA}$ for the 2006-2009 unusual minimum.

\end{abstract}


\section{Introduction} \label{sec1}

The heliospheric conditions were quite unusual during the solar minimum period 2006-2009, between Solar Cycle 23/24. The minimum was unusually long and deep, the heliospheric magnetic field (HMF) was much weaker, and the solar wind was slowest and least dense compared to previous Solar Cycles. However, the tilt angle ($\alpha$) of the heliospheric current sheet (HCS) was more warped (less flat) and had not decreased rapidly, as compared to previous minima, but reached a minimum value at the end of 2009 (Aslam \& Badruddin 2012; Potgieter et al. 2015). The very low and almost constant intensity of the solar polar field (about half of the previous two minima) during this unusual minimum activity period was also of interest from the point of view of cosmic ray (CR) modulation (e.g. Wang et al. 2009; Jian et al. 2011). A record high level of CR intensity, since the beginning of the era of neutron monitor (NM) observations, was observed by these ground-based detectors, and the highest ever galactic proton spectrum at Earth was observed by $\it{PAMELA}$ during the end of 2009; see the discussion by Heber et al. (2009), Mewaldt et al. (2010), Potgieter et al. (2014), Aslam \& Badruddin (2015), and other relevant references there-in. This unusual increase in CR intensity also indicates that this minimum had an extraordinary effect on the properties of the magnetic structure shielding the Earth to allow such an increase in CRs (White et al. 2011). 

Through the level of solar activity, the velocity of the solar wind, the tilt angle of the HCS, and the magnitude of the HMF, the Sun controls the heliospheric structure and the modulation of CRs (as an example, see McDonald et al. 2010; Vos \& Potgieter 2015, 2016). The Sun’s magnetic field polarity reverses during ever solar  maximum activity phase; if the field lines are outward directed from the Sun’s northern hemisphere, it is defined as positive polarity (A$>$0) and if the field lines are outward from the Sun’s southern hemisphere, it is defined as negative polarity (A$<$0). The HCS separates this two oppositely directed magnetic polarity hemispheres. As $\alpha$ varies with solar activity, the changing HCS has a significant effect on CR modulation through the drift motion of CR particles (e.g. Jokipii \& Thomas 1981; Potgieter \& Moraal 1985; Ferreira et al. 2003; Zhao et al. 2014). The period from 2001 to May 2012 (including the unusual 2006-2009 minimum) had a negative magnetic polarity (A$<$0) (see http://wso.stanford.edu).

CRs are subjected to four distinct transport processes in the solar wind plasma flow with its imbedded HMF, a) Convection because of the outward directed solar wind velocity, b) Gradient, curvature and current sheet drifts, c) Adiabatic energy changes depending on the sign of the divergence of the solar wind velocity, and d) Spatial diffusion caused by the scattering off random magnetic irregularities (for reviews, see e.g. Heber 2013; Potgieter 2013; Kota 2013). The drift process has different effects in each solar activity cycle, however, the diffusion and convection processes are solar activity cycle dependent rather than on the solar magnetic polarity.

According to drift models, the positively charged CR particles (protons, positrons, helium, etc.) drift inward mainly along the heliospheric equatoral regions and outward via the polar regions of the heliosphere during the A$<$ 0 polarity phase. On the other hand, negatively charged CR particles (electrons, anti-protons) will drift downward from the poles and outward through the equatorial regions during this polarity phase. The particle drift direction reverse for both positively and negatively charged CR particles during the opposite polarity configuration (A$>$0) (Kota \& Jokipii 1983). This means that when Galactic positrons drift inwards through the equatorial regions, the wavy HCS plays a prominent role on their modulation during the A$<$0 polarity phase, as happened during the minimum of 2006-2009.

The detailed study and illustration of the response of Galactic protons during this unusual solar minimum period using a comprehensive three-dimensional (3D) heliospheric modulation model including gradient, curvature, and HCS drift, applied to the $\it{PAMELA}$ proton observations (Adriani et al. 2013a), was carried out by Potgieter et al. (2014) and Vos \& Potgieter (2015). Similar detailed heliospheric modulation of Galactic electrons was also studied by Potgieter et al. (2015) and Potgieter \& Vos (2017) by utilizing the $\it{PAMELA}$ electron observations (Adriani et al. 2015; Munini et al. 2015) over this period. These authors concluded that all four modulation processes played important roles, but diffusion was relatively dominant during this unusual minimum in contrast to the  paradigm for the modulation of CRs that drift should be dominating during solar minima, but that diffusion is considered to become the dominant process during solar maximum. The current study is a continuation of these studies, now focusing on the Galactic positron observations made by $\it{PAMELA}$ from mid-2006 to the end of 2009 (Adriani et al. 2013b; Munini 2015; Munini et al. 2017). Such a comprehensive study for positrons has not been done before.

\section{The PAMELA Positron Observations from July 2006- December 2009} \label{sec2}

The $\it{PAMELA}$ detector is a space-borne particle spectrometer, specially designed to measure the spectra of primary and secondary components of CRs at Earth (Adriani et al. 2014, 2017; Boezio et al.2017). The main focus was the indirect study of dark matter through the detection of the anti-proton and positron spectra up to 200 GeV. The Carrington rotation averaged proton spectra from $\it{PAMELA}$ for the period 2006-2009, over an energy range of $\approx$ 80 MeV to 50 GeV, was presented by Adriani et al. (2013a). The $\it{PAMELA}$ electron spectra for the same period over the energy range $\approx$ 80 MeV to 40 GeV is six-month averaged, instead of Carrington rotation averaged, were also reported by Adriani et al. (2015). The first positron spectra above 1 GeV from $\it{PAMELA}$ were published by Adriani et al. (2013b). Later, the six-month averaged positron spectra for the period July 2006 to end of 2009 down to 80 MeV were reported by Munini (2015), with the abundance of the positron intensity roughly $10\%$ of the electron intensity and around $1\%$ of the proton intensity at 10 GeV.
\begin{figure}[!htp]
\plotone{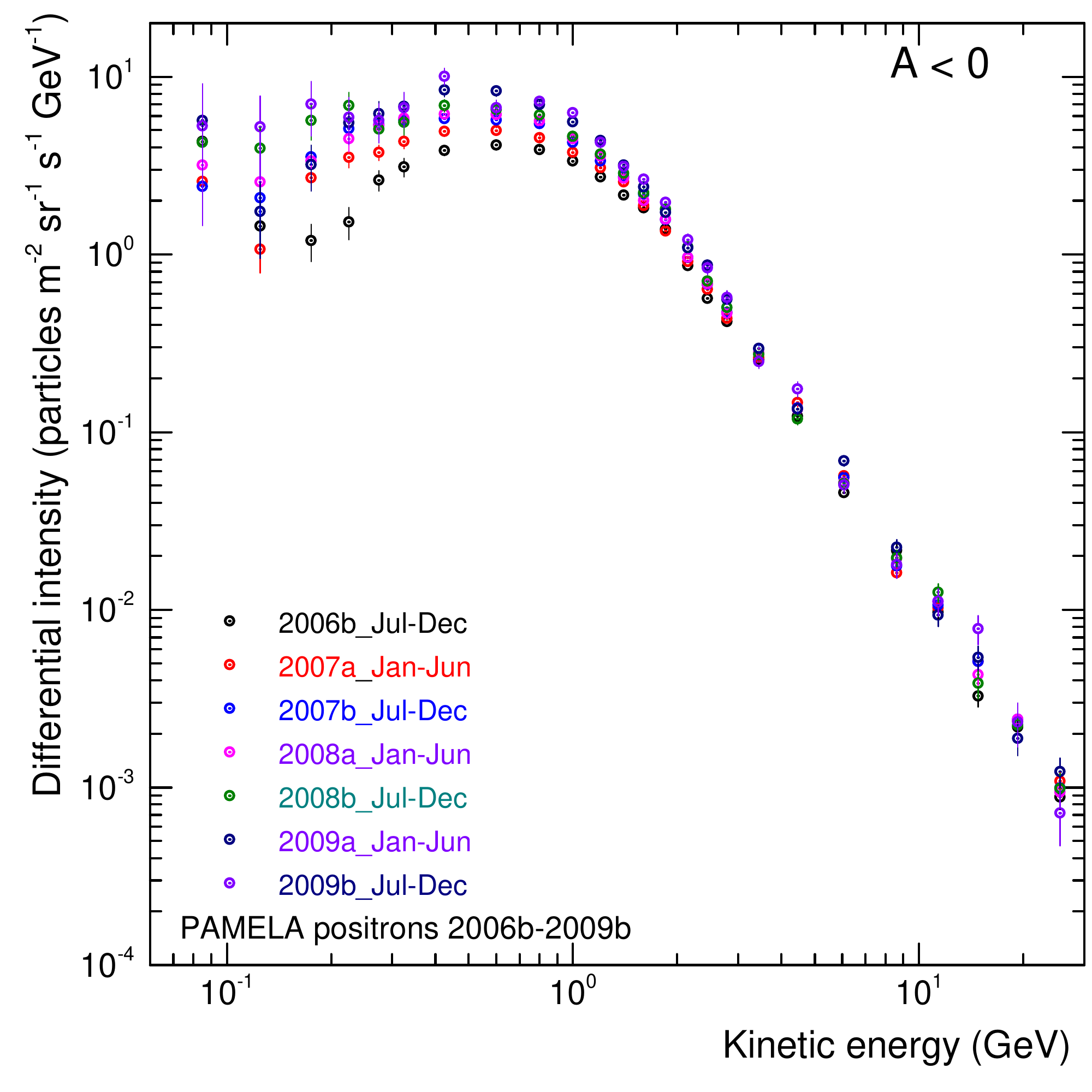}
\caption{Positron energy spectra observed by the $\it{PAMELA}$ particle spectrometer, as six-month averages for seven periods starting from 2006b (July-December 2006) up to 2009b (July-December 2009); adapted from Munini (2015). The spectrum labeled 2009b was the highest intensity observed. The confidence level of the spectra below 200 MeV is relatively weaker as indicated by the larger error bars. \label{fig1}}
\end{figure}

Figure 1 shows the positron energy spectra measured by $\it{PAMELA}$, averaged over six-month periods, from July 2006 to December 2009 (adapted from Munini 2015), The total period is divided into 7 semesters of six-months, and labeled \enquote{a}: January-June, and \enquote{b}: July-December (i.e. 2006b, second semester of the year) represents the average over July- December 2006, and 2007a (first semester of the year) represents the average over January-June 2007). The positron spectra presented here have an energy range from $\approx$ 80 MeV to 30 GeV, it should be noted that the flux data below 200 MeV are affected by significant systematic uncertainties. Evidently, the highest intensity was observed during the 2009b period. 
\begin{figure}[!htp]
\plotone{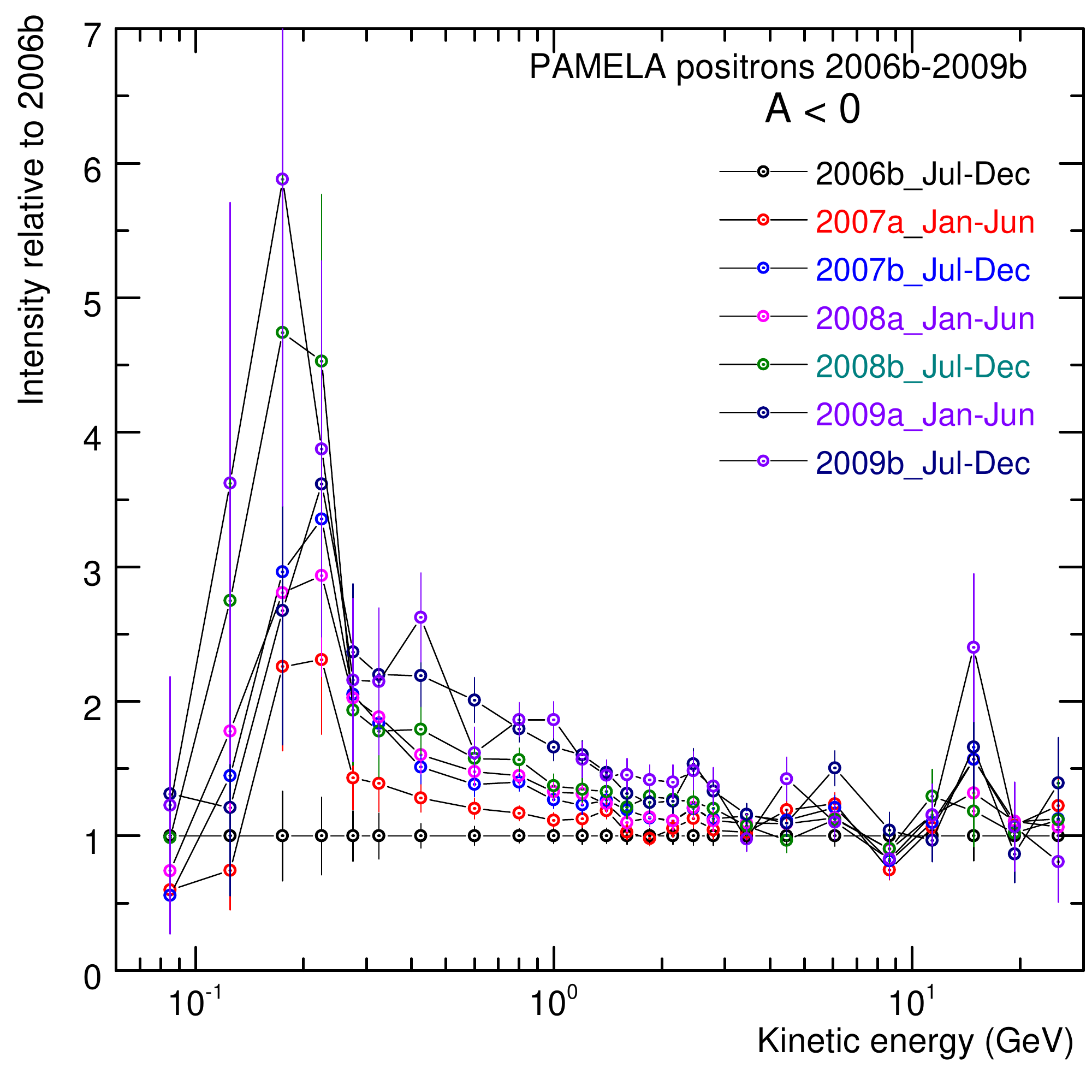}
\caption{Consecutive ratios of the $\it{PAMELA}$ positron energy spectra shown in the Figure 1, relative to that for 2006b as a function of kinetic energy. The spread between the 2006b and 2009b spectra represents modulation-dependent changes over the considered time period, July 2006 to December 2009. \label{fig2}}
\end{figure}

Figure 2 shows the positron intensity ratio relative to that for 2006b, as a function of kinetic energy in GeV. Note that the spectra up to 200 MeV is the most responsive to changes in modulation conditions and have undergone an increase of a factor of $\approx$ 6. This increase comes down to a factor of $\approx$ 2 at 1 GeV and become less pronounced above 10 GeV as the solar modulation of CRs subsides; above $\approx$ 30 GeV the ratio shows very little changes (see also Strauss \& Potgieter 2014).

The prime objective of this work is to reproduce the $\it{PAMELA}$ positron spectra observed during the period July 2006-December 2009, using a comprehensive three-dimensional modulation model which is described below.

\section{The Very Local Interstellar Spectrum for Positrons} \label{sec3}

As an initial condition a Galactic positron spectrum, more specifically a local interstellar spectrum (LIS), has to be specified in the model to be used as the input spectrum which then is modulated from a given heliospheric boundary up to the Earth at 1 au. 
\begin{figure}[!htp]
\plotone{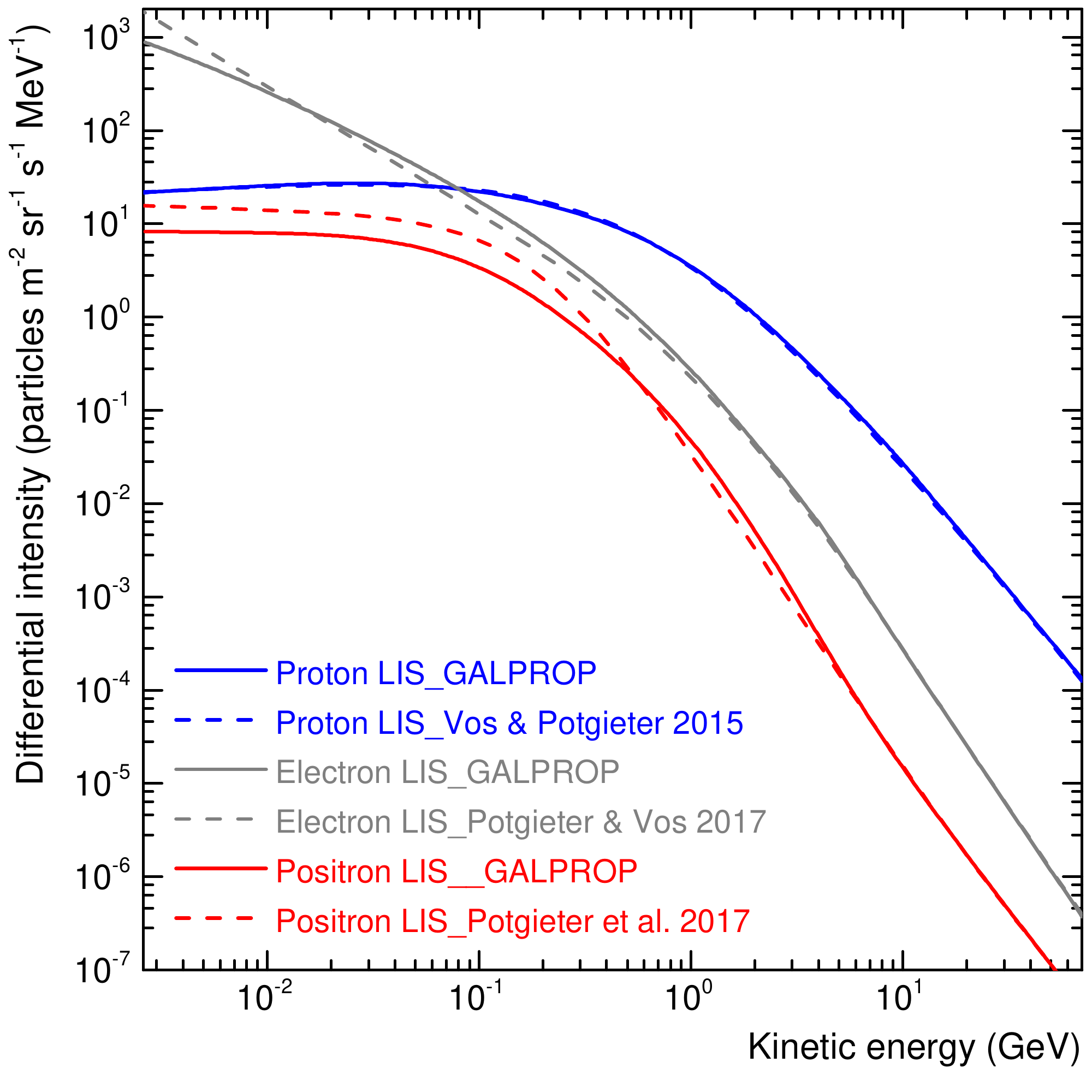}
\caption{Solid lines depict LIS for CR protons, electrons and positrons as computed with GALPROP; the blue dashed line shows the proton LIS constructed and used by Vos \& Potgieter (2015); grey dashed line shows the electron LIS constructed and used by Potgieter \& Vos (2017), and the red dashed line shows the positron LIS used by Potgieter et al. (2017). \label{fig3}}
\end{figure}

Figure 3 illustrates the differences (variations) between the LIS of protons, electrons and positrons computed with the web-version of the GALPROP code (see e.g. Moskalenko \& Strong 1998) and the corresponding LIS from previous modulation studies: The proton LIS was constructed by Vos \& Potgieter (2015) and is similar to the GALPROP proton LIS; the electron LIS constructed by Potgieter et al. (2015) and Potgieter \& Vos (2017) and the positron LIS constructed by Potgieter et al. (2017) show however deviations as a function of kinetic energy from the LIS produced by the GALPROP code. These authors used $\it{PAMELA}$ and later also $\it{AMS}$-02 observations at the Earth at high energies where modulation is negligible, and particularly Voyager 1 observations (Stone et al. 2013; Gurnett et al. 2013; Webber \& McDonald 2013) at very low energies beyond the heliopause to construct the presented LIS of protons and electrons; and $\it{PAMELA}$ 2006-2009, also $\it{AMS}$-02 2011-2013 observations to construct the positron LIS.

For our modelling, we first revisited electron modulation and reproduced the $\it{PAMELA}$ electron spectra for seven semesters from 2006b to 2009b, similar to what was done by Potgieter et al. (2015). For this study we used the electron LIS computed with the GALPROP code. We use this approach to verify our modulation approach and to constrain all the modulation parameters in order to repeat the same process for positrons. Arguing that the only two differences between electron and positron modulation are their respective LIS's and the particle drift that they experience, we set out to reproduce the $\it{PAMELA}$ positron spectra at the Earth using the LIS for positrons as computed with GALPROP. However, it soon became evident that we could not reproduce the modulated $\it{PAMELA}$ positron spectra when using this LIS, despite various tuning of the the GALPROP code, so we had to construct empirically a different LIS but still based on the GALPROP LIS. In Figure 4 we illustrate the differences between our LIS and the one from GALPROP (black lines) as a function of kinetic energy and the corresponding differences between the subsequent modulated spectra for the periods 2006b (blue lines) and 2009b (red lines) only. Evidently, the differences are meaningful. This modified LIS for positrons is used in the rest of our study and could reproduce the observed positron spectra as will be shown and discussed below. The tuning of the GALPROP code as mentioned above is described in detail by Bisschoff (2017); see also Bisschoff \& Potgieter (2014, 2016) and Bisschoff et al. (2018).
\begin{figure}[!htp]
\plotone{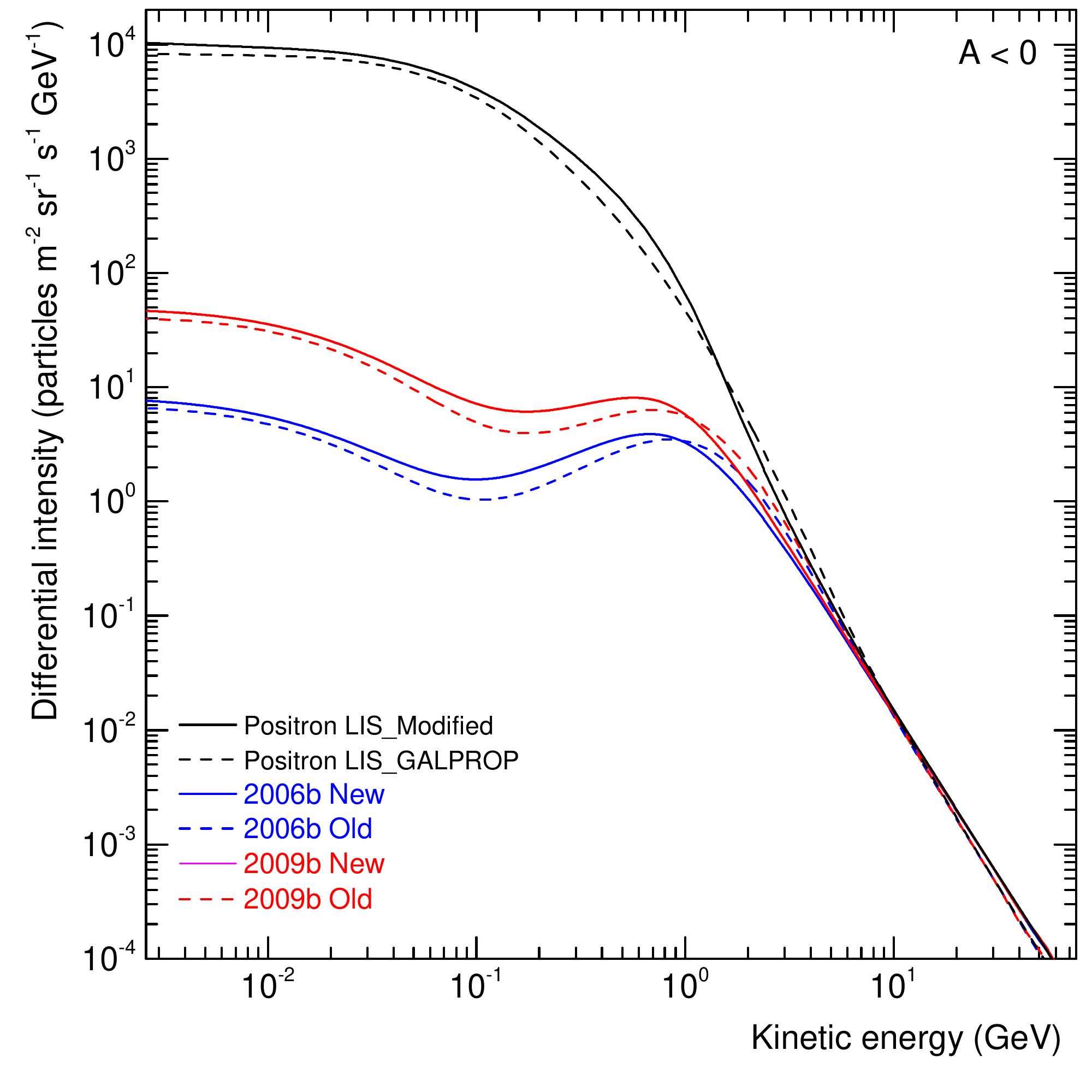}
\caption{A modified LIS (black solid line) for Galactic positrons with its corresponding modulated spectra at the Earth for the six-month periods 2006b (solid blue line) and 2009b (solid red lines) in comparison with the positron LIS (black dashed line) computed with GALPROP and its corresponding modulated spectra for 2006b (blue dashed line) and 2009b (red dashed line).\label{fig4}}
\end{figure}

\section{Numerical Simulation of PAMELA Positron Spectra} \label{sec4}

A full three-dimensional (3D) model, which is based on the numerical solution of Parker’s heliospheric transport equation (TPE; Parker 1965), is used to compute the differential intensity of CR positrons over an energy range from 1 MeV to 70 GeV at the Earth (1 au) and at different radial distances up to 122 au, the position of heliopause in the model where the LIS is specified as an initial condition. No modulation is considered beyond the heliopause; for such an approach, see Luo et al. (2015).

Parker's transport equation is described as:  
\begin{equation}
\frac{\partial f}{\partial t} = - \vec{V}_{sw} \cdot \nabla \it{f} - \langle \vec {v}_{D} \rangle \cdot \nabla \it{f} + \nabla \cdot (\bf{K}_{s} \cdot \nabla \it{f}) + \frac {1}{3} (\nabla \cdot \vec{V}_{sw}) \frac {\partial f} {\partial  ln  P}, \label{Eq1}
\end{equation}

where $f (\vec {r}, P, t)$ is the CR distribution function, $\it{P}$ is rigidity, $\it{t}$ is time, and $\vec {r}$ is the vector position in 3D, with the three coordinates $\it{r}$, $\theta$, and $\phi$ specified in a heliocentric spherical coordinate system where the equatorial plane is at a polar angle of $\theta = 90^{\circ}$. The four terms shown on the right-hand side of the Equation (\ref {Eq1}) represent the four major physical processes which CR particles undergo when they enter and travel through the heliosphere up to the Earth.    

The first term represents the outward convection caused by the expanding solar wind with velocity ($\vec {V}_{sw}$), the second term represents the averaged particle drift velocity $\langle \vec {v}_{D} \rangle$ (pitch angle averaged guiding center drift velocity), which is described by 
\begin{equation}
\langle \vec {v}_{D} \rangle = \nabla \times K_{D} \frac{\vec B}{B}
\end{equation}
where $K_{D}$ is the generalized drift coefficient, and $\vec{B}$ is the heliospheric magnetic field (HMF) vector with magnitude $\it{B}$. Both the geometry and magnitude of the HMF are important for the  modulation process, particularly in the case of gradient, curvature and HCS drift. As a departure point, a straight-forward Parker HMF (Parker 1958) is assumed: 
\begin{equation}
\vec {B} = B_{0}A\Bigg[\frac{r_{0}}{r}\Bigg]^{2}(\hat {e}_{r}-tan\psi \hat{e}_{\phi})[1-2H(\theta - \theta^{'})],
\end{equation}     

where $\hat {e}_{r}$ and $\hat {e}_{\phi}$ are unit vectors in the radial and azimuthal directions, $A$ = $\pm$1, expresses the polarity phase of the Sun ($A$ = +1 is $A>$0; positive polarity and $A$ = -1 is $A<$0; negative polarity); $B_{0}$ is the magnitude of the HMF at Earth (i.e. $r_{0}$ = 1 au) and the spiral angle $\psi$ is the angle between the radial direction and the HMF field line at any given position defined by:
\begin{equation}
  \it{tan}  \psi = \Omega \frac {(r - r_{\odot})} {V_{sw}} sin\theta, 
\end{equation}     

where $r_{\odot}$ is the solar radius (0.005 au) and $\Omega$ is the average angular rotation speed of the Sun (2.66 $\times$ 10$^{-6}$ rad s$^{-1}$). The magnitude of this Parker HMF is 
\begin{equation}
B = B_{0}\Bigg[\frac{r_{0}}{r}\Bigg]^{2}\sqrt{1 + (tan \psi)^{2}.}
\end{equation} 

Smith \&  Bieber (1991) suggested a modification to the Parker field based on their observation that the magnetic field spirals are relatively more tightly wound than that predicted by the original Parker theory. They argued that the differential rotation of the Sun would cause small azimuthal magnetic field components to develop and that would lead to larger spiral angles at larger radial distances. They proposed a modification so that the expression for the HMF spiral angle (Equation 4) becomes:
\begin{equation}
tan \psi = \frac {\Omega (r - r_{b}) sin\theta} {V_{sw} (r, \theta)} - \frac {r V_{sw} (r_{b}, \theta)} {r_{b} V_{sw} (r, \theta)} \Bigg( \frac {B_{T}(r_{b})}{B_{R}(r_{b})}\Bigg)
\end{equation}
 
where $B_{T}(r_{b})/ B_{R}(r_{b})$ is the ratio of the azimuthal to the radial magnetic field components at a position $r_{b}$ near the solar surface. Smith \& Bieber (1991) showed that the ratio $B_{T}(r_{b})/ B_{R}(r_{b})$ $\approx$ -0.02 at a position $r_{b}$ = 20$r_{\odot}$. With $r_{\odot}$ = 0.005 au as the solar radius, the value $r_{b}$ = 20$r_{\odot}$ and the ratio $B_{T} (r_{b})/B_{R} (r_{b})$ = -0.02 are constants that determine the HMF modification. This modification has a significant effect on the HMF structure at high latitude regions; it keeps the basic Parkerian geometry but modifies its magnitude progressively toward the poles of the heliosphere. The motivation for this modification from a modulation modelling point of view was discussed in detail by Potgieter (2013) and Potgieter et al. (2015), while Raath et al. (2015) gave an elaborate discussion of its relevance and of its CR modulation effects. 

For a modified Parker-type HMF ($\vec B_{m}$), such as the Smith-Bieber modification, with magnitude ($B_{m}$), the expression for a generalized drift coefficient is:
\begin{equation} 
K_{D} = \frac {\beta P} {3B_{m}} f_{D} = \frac {\beta P} {3B_{m}} \Bigg[ \frac {(\omega \tau)^{2}}{1+(\omega \tau)^{2}},\Bigg]
\end{equation}

where $\beta$ = $v/ c$ is the ratio of particle speed to the speed of light, and $\omega$ is the particle gyro-frequency with $\tau$ the average time between the scattering of CR particles in the HMF. In most of the numerical modeling studies, it is assumed that $\omega \tau \gg 1$ in the heliosphere so that the drift coefficient takes its simplest form:
\begin{equation} 
K_{D} = \frac {\beta P}{3B_{m}},
\end{equation}

which is known as weak scattering drift. Establishing $\tau$ is complicated and controversial; an elaborate turbulence theory is required to understand how $\omega\tau$ should change as a function of rigidity and space throughout the entire heliosphere; see also the discussion by Ngobeni \& Potgieter (2015). The term $f_{D}$ is called the drift reduction factor and is determined by how diffusive scattering is described; if $f_{D}$ = 0, then $K_{D}$ and therefore $\langle v_{D} \rangle$ become zero, meaning drift effects will vanish from the modulation model; if $f_{D}$ = 1, drift is at a maximum, so that $K_{D}$ will have the weak scattering value. Then, drift effects on CR modulation are very large and dominant as originally applied in numerical models by Jokipii and Thomas (1981) and Kota and Jokipii (1983).  

This function can be used to adjust the rigidity dependence of $K_{D}$, which is the most effective direct way of suppressing drift effects at low rigidities, so that Equation (7) becomes:
\begin{equation}
K_{D} = \frac {\beta P} {3B_{m}} f_{D} = K_{A0}  \frac {\beta P} {3B_{m}} \frac {(P/P_{A0})^{2}}{1+(P/P_{A0})^{2}}.
\end{equation}           

Here $K_{A0}$ is a dimensionless constant that could be ranging from 0 to 1.0; if $K_{A0}$ = 1.0, it is called 100\% drift (full weak scattering). In this study, we keep $K_{A0}$ = 0.90 and $P_{A0}$ = 0.90 GV for all the seven semesters (mid-2006 to end of 2009), which means particle drift is at a 90\% level during this unusual solar minimum period, but below $\approx$ 1.0 GV, the drift coefficient is reduced with respect to the weak scattering approach in Equation (8). For more details, see also Potgieter (2013, 2014) and Nndanganeni \& Potgieter (2016). 

The drift scale ($\lambda_{A}$) with respect to rigidity at the Earth (1 au with polar angle $\theta$ = 90$^{\circ}$) is shown in Figure 6. The fast decrease in $\lambda_{A}$ below rigidity $\approx$ 1 GV is the direct result of the scaling of $f_{D}$. The weak scattering drift (Equation 8) and modified drift coefficient (Equation 9) have the same rigidity dependence above $\approx$ 1 GV. The reason why this is required in numerical modeling was discussed in detail in the review by Heber \& Potgieter 2006; see also Potgieter (2014). However, in the case of low energy positrons and electrons, particle drift becomes negligible because their transport in the heliosphere is dominated by the diffusion process; see the review by Potgieter (2017). Unlike CR protons and heavy nuclei, the electrons and positrons experience far less adiabatic energy losses at low energies, so that they respond directly to changes of the diffusion coefficients down to very low kinetic energy. Previous numerical studies of Galactic protons and electrons by Potgieter et al. (2015) and Potgieter \& Vos (2017) used $K_{A0}$ = 1.0 and $P_{A0}$ = 0.55 GV for the 2006-2009 minimum period.

Reducing drifts using $B$ and $f_{D}$ is the way of explicit drift reduction, whereas implicit drift reduction is a way of reducing drift effect without changing $K_{D}$. This is made possible by reducing the diffusive process by increasing any of the appropriate diffusion coefficients of the diffusion tensor $\bf{K}_{s}$. This was illustrated in detail for CR electrons by Nndanganeni \& Potgieter (2016).
       
The third term in the right hand side of Equation (1) indicates the spatial diffusion caused by the scattering of CRs, where $\bf{K}_{s}$ is the symmetry diffusion tensor, and the last term represents the adiabatic energy change, which depends on the sign of the divergence of the  $\vec {V}_{sw}$. If $(\nabla \cdot \vec{V}_{sw})>0 $ adiabatic energy losses occur as is the case in most of the heliosphere, except inside the heliosheath where we assume that $(\nabla \cdot \vec{V}_{sw}) = 0$.  Adiabatic energy loss is one of the major mechanisms and is important for CR modulation in most of the heliosphere, but far less dominant in the case of electrons and positrons than for protons and other heavy CR species; see also the illustrations by e.g. Moraal and Potgieter (1982).

When we write the TPE given by Equation (1) in heliocentric spherical coordinate system, there are nine elements of the diffusion tensor $\bf{K}_{s}$: $K_{rr}$, $K_{r\theta}$, $K_{r\phi}$, $K_{\theta r}$, $K_{\theta \theta}$, $K_{\theta \phi}$, $K_{\phi r}$, $K_{\phi \theta}$, and $K_{\phi \phi}$, based on a Parker-type HMF with a radial solar wind speed $V_{sw}$. Here, $K_{rr}$, $K_{r\phi}$, $K_{\theta \theta}$, $K_{\phi r}$, and $K_{\phi \phi}$ describes the diffusion process and $K_{r\theta}$, $K_{\theta r}$, $K_{\theta \phi}$, and $K_{\phi \theta}$ describes the gradient, curvature and HCS drift. These diffusion coefficients can be expressed in a 3D heliocentric spherical coordinate system as follows: 
\begin{equation}        
K_{rr} = K_{\parallel} \it{cos}^{2} \psi + K_{\perp r} \it{sin}^{2} \psi. 
\end{equation}
\begin{equation} 
K_{\theta \theta} = K_{\perp \theta}. 
\end{equation}
\begin{equation} 
K_{\phi \phi} = K_{\perp r} \it{cos}^{2} \psi + K_{\parallel} \it{sin}^{2} \psi. 
\end{equation}
\begin{equation} 
K_{\phi r} = (K_{\perp r} - K_{\parallel}) \it{cos} \psi \it{sin} \psi = K_{r \phi}. 
\end{equation}

The effective radial diffusion coefficient $K_{rr}$ is a combination of the parallel diffusion coefficient ($K_{\parallel}$) and the radial perpendicular diffusion coefficient ($K_{\perp r}$), with $\psi$ the spiral angle; $K_{\theta \theta}$ = $K_{\perp \theta}$ is the effective perpendicular diffusion coefficient in the polar direction. $K_{\phi \phi}$ and $K_{\phi r}$ describe the effective diffusion in the azimuthal direction and diffusion coefficient in the $\phi r$-plane respectively. 

Potgieter et al. (2014, 2015) described the 3D model used for this study in detail, with recent, comprehensive reviews of the underlying theory for the global modulation of CRs during a quite heliosphere given by Potgieter (2013, 2014, 2017).  

\subsection{Calculation and Selection of Modulation parameters} \label{sec4.1}

The 3D numerical model for this modulation study is for a steady-state, so that for the left side of Equation (1), $\partial f /\partial t = 0$; this means that transient events such as Frobush decreases, cannot be studied with this model. Determination of the averaged values for time-varying modulation parameters is an important part of the modeling method when working with such a model, e.g. the tilt angle $\alpha$ of the HCS and magnitude of the HMF at Earth $B_{e}$ are varying continuously according to the solar activity. The time variations in $\alpha$ and $B_{e}$ are included in the numerical model to set up realistic modulation conditions which is reasonable for the relatively quiet state of the heliosphere.
 
\begin{figure}[!htp]
\plotone{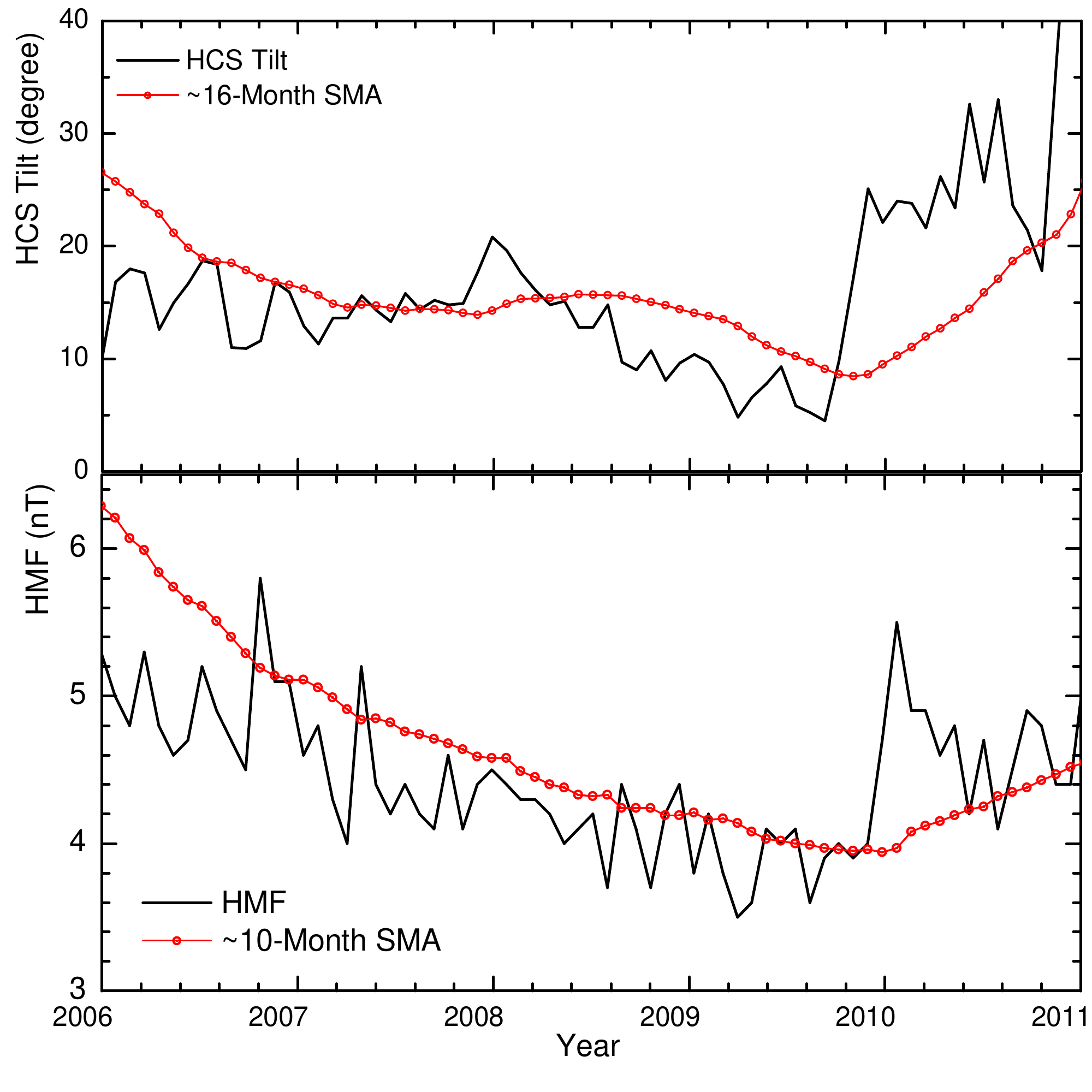}
\caption{Top Panel: Tilt angle $\alpha$ of the HCS (black line) at the Earth from 2006 to 2011, taken from http://wso.stanford.edu, along with 17 Carrington rotation ($\approx$ 15-months) moving averages (red line). Bottom Panel: Magnitude of the HMF (black line) at the Earth ($B_{e}$) for the same period taken from http://omniweb.gsfc.nasa.gov, along with $\approx$ 10-months moving averages (red line). The averaged values used for each semester are shown in Table 1. \label{fig5}}
\end{figure}

The average solar wind speed in the slow solar wind region is assumed $\approx430$ kms$^{-1}$ and $\approx750$ kms$^{-1}$ in the fast solar wind region. During minimum to moderate solar activity periods the HCS is mostly confined to the ecliptic region (within $\approx$ 30$^{\circ}$), remaining in the slow solar wind region. The slow solar wind velocity thus is considered to calculate the time it takes changes in $\alpha$ to travel from the Sun to the heliopause. However, the HMF is not confined to the elliptic region, so the weighted average of both slow and fast solar wind velocities is used to calculate the time taken to reach the heliopause from the Sun. The propagation time calculated for $\alpha$ is $\approx$16-months (17-solar rotations) and for the HMF it is $\approx$10-months (11-solar rotations). The moving average over this propagation time for both $\alpha$ and $B_{e}$ are used as intrinsic parameters in this modeling approach, similar to what Potgieter et al. (2015) and Vos \& Potgieter (2015) assumed. The observed $\alpha$ and $B_{e}$ are shown in Figure \ref {fig5}, together with this averaging process, and is summarized in Table 1. 

Richardson \& Wang (2011) reported that because of the dynamic nature of the heliosphere, the position of the termination shock (TS) also varies with solar activity. So in addition to solar activity related variables mentioned above, the changing position of TS is also incorporated in the model since it affects the modulated intensities at the Earth (see also e.g. Manuel et al. 2014). In the model, the position of TS is changed from 88 au in 2006 to 86 au in 2007, 84 au in 2008 and 80 au in 2009 following Potgieter et al. (2015).  The calculated moving averages over the propagation time of $\alpha$ and $B_{e}$ for each semester along with the position of TS is tabulated in Table \ref {Table1}. Note that the value of $\alpha$ was less than 10$^{\circ}$ during the 2009b period (see Figure \ref {fig5}).

The general expression for the diffusion coefficient parallel to the average background HMF is given by:
\begin{equation}
K_{\parallel} = (K_{\parallel})_{0} \beta \Bigg(\frac {B_{0}}{B}\Bigg) \Bigg(\frac {P}{P_{0}}\Bigg)^{c_{1}} \left[ \frac {\Bigg(\frac {P}{P_{0}}\Bigg)^{c_{3}} + \Bigg(\frac {P_{k}}{P_{0}}\Bigg)^{c_{3}}}{ 1+ \Bigg(\frac {P_{k}}{P_{0}}\Bigg)^{c_{3}}} \right]^{\frac {c_{2 \parallel} - c_{1}}{c_{3}}},
\end{equation}

with $(K_{\parallel})_{0}$ is a scaling constant in units of $10^{22}$ cm$^{2}$s$^{-1}$ , with the rest of the equation written to be dimensionless with $P_{0}$ = 1 GV, and $B_{0}$ = 1 nT (in order to keep the same cm$^{2}$ s$^{-1}$  unit). Here $c_{1}$ is a power index that may change with time if required; $c_{2 \parallel}$ and $c_{2 \perp}$ ($c_{2 \perp}$ = 0.75 $c_{2 \parallel}$) together with $c_{1}$ determine the slope of the rigidity dependence, respectively, above and below a rigidity with the value $P_{k}$ which may change with time if required, see Table 1; $c_{3}$ determines the smoothness of the transition. The rigidity dependence of $K_{\parallel}$ is thus a combination of two power laws; $P_{K}$ determines the rigidity where the break in the power law occurs and the value of $c_{1}$ determines the slope of the power law at rigidities below $P_{k}$. 
The radial dependence of the diffusion coefficients in the inner heliosphere (less than 5 au) was adjusted according to Vos \& Potgieter (2015, 2016) in order to reproduce the radial gradients as observed by Ulysses (Gieseler \& Heber, 2016) which requires an increase in the diffusion coefficients at these radial distances.
\begin{figure}[!htp]
\plotone{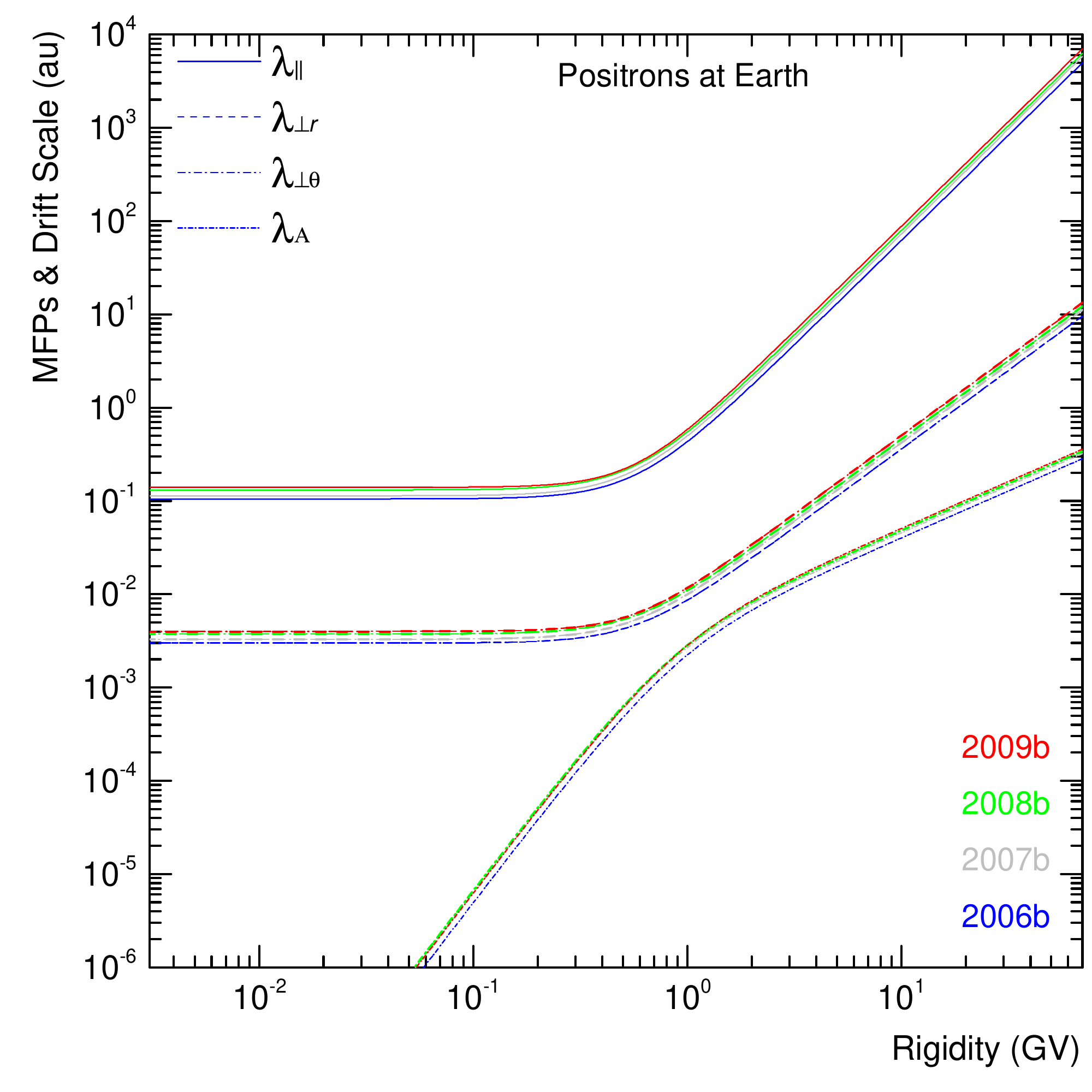}
\caption{Rigidity dependence of the positron MFPs (both for diffusion parallel and perpendicular to the magnetic field lines) and the drift scale at Earth. The rigidity dependence of $\lambda_{\perp r}$ and $\lambda_{\perp \theta}$ are identical. \label{fig6}}
\end{figure}

Perpendicular diffusion in the radial direction is assumed to scale spatially similar to Equation (14) but with a different rigidity dependence at higher rigidities, in the case of positrons and electrons the value of c$_{1}$ = 0.0, so the term  $(P/P_{0})^{c_{1}}$ will be unity, so that,  
\begin{equation}
K_{\perp r} = 0.02 (K_{\parallel})_{0}  \beta  \Bigg(\frac {B_{0}}{B}\Bigg) \left[ \frac {\Bigg(\frac {P}{P_{0}}\Bigg)^{c_{3}} + \Bigg(\frac {P_{k}}{P_{0}}\Bigg)^{c_{3}}} { 1+ \Bigg(\frac {P_{k}}{P_{0}}\Bigg)^{c_{3}}} \right]^{\frac {c_{2 \perp}}{c_{3}}}.
\end{equation}
It means, $K_{\perp r}$ = 0.02 $K_{\parallel}$, is a straightforward, reasonable and widely used assumption. 

On the other hand, the polar perpendicular diffusion ($K_{\perp \theta}$) is anisotropic in the inner heliosphere, with consensus that $K_{\perp \theta}$ $>$ $K_{\perp r}$ away from the equatorial regions as discussed by Potgieter (2000) and Potgieter et al. (2014).
 
The perpendicular diffusion coefficient in the polar direction is given by
\begin{equation}
K_{\perp \theta} = 0.02 K_{\parallel} f_{\perp \theta} = K_{\perp r} f_{\perp \theta} 
\end{equation}
with
\begin{equation}
f_{\perp \theta} = A^{+} \mp A^{-} tanh[8(\theta_{A} - 90^{\circ}) \pm \theta_{F}].
\end{equation}

Here, A$^{\pm} = (d_{\perp \theta} \pm 1)/2 $, $\theta_{F}$ = 35$^{\circ}$, $\theta_{A}$ = $\theta$ for $\theta \leq 90^{\circ}$ but  $\theta_{A}$ = 180$^{\circ}$ - $\theta$ with $\theta \geq$  90$^{\circ}$ and $d_{\perp \theta}$ = 6.0 (in this study, see Table 1). 
This means that the polar perpendicular diffusion K$_{\perp \theta}$ can be enhanced towards the heliospheric poles by a factor d$_{\perp \theta}$; we assumed d$_{\perp \theta}$ = 6 with respect to the value K$_{\parallel}$ in the equatorial region of the heliosphere. For a motivation of this approach, see Potgieter (2000), Ferreira et al. (2003), Ngobeni \& Potgieter (2011) and Potgieter et al. (2015).

The modulation parameter values required to reproduce the PAMELA positron spectra, discussed next, for the period 2006b to 2009b are summarized in Table \ref {Table1}. The perpendicular diffusion in the radial and polar directions are represented by $K^{0}_{\perp r}$ and  $K^{0}_{\perp \theta}$ in Table 1. The rigidity dependence of the diffusion coefficients and drift coefficient at the Earth are illustrated together in Figure \ref{fig6}. The diffusion coefficients ($K$) are related to corresponding mean free paths (MFPs) ($\lambda$) by, $K$ = $\lambda$($v/3$), where $v$ is the speed of positrons. Figure \ref {fig6} shows the rigidity dependence of $\lambda_{\parallel}$ and $\lambda_{\perp} \equiv \lambda_{\perp r} \equiv \lambda_{\perp \theta}$  as well as the drift scale ($\lambda_{A}$) at Earth.  We have attempted to keep most of these variables the same for all the seven semesters (2006b - 2009b), except $\lambda_{\parallel}$, $c_{3}$ and $P_{k}$, as shown in Table 1. 

\subsection{Comparison of modeling results with observations} \label{sec4.2}
The Parker TPE is solved for each semester (2006b to 2009b) by using the calculated average values of the HCS tilt and the HMF magnitude at the Earth, and by adjusting the diffusion and drift coefficients. We successfully reproduced the six-month averaged $\it{PAMELA}$ positron spectra from July 2006 to December 2009, by carefully adjusting the diffusion and drift coefficients.

\begin{figure}[!htp]
\plotone{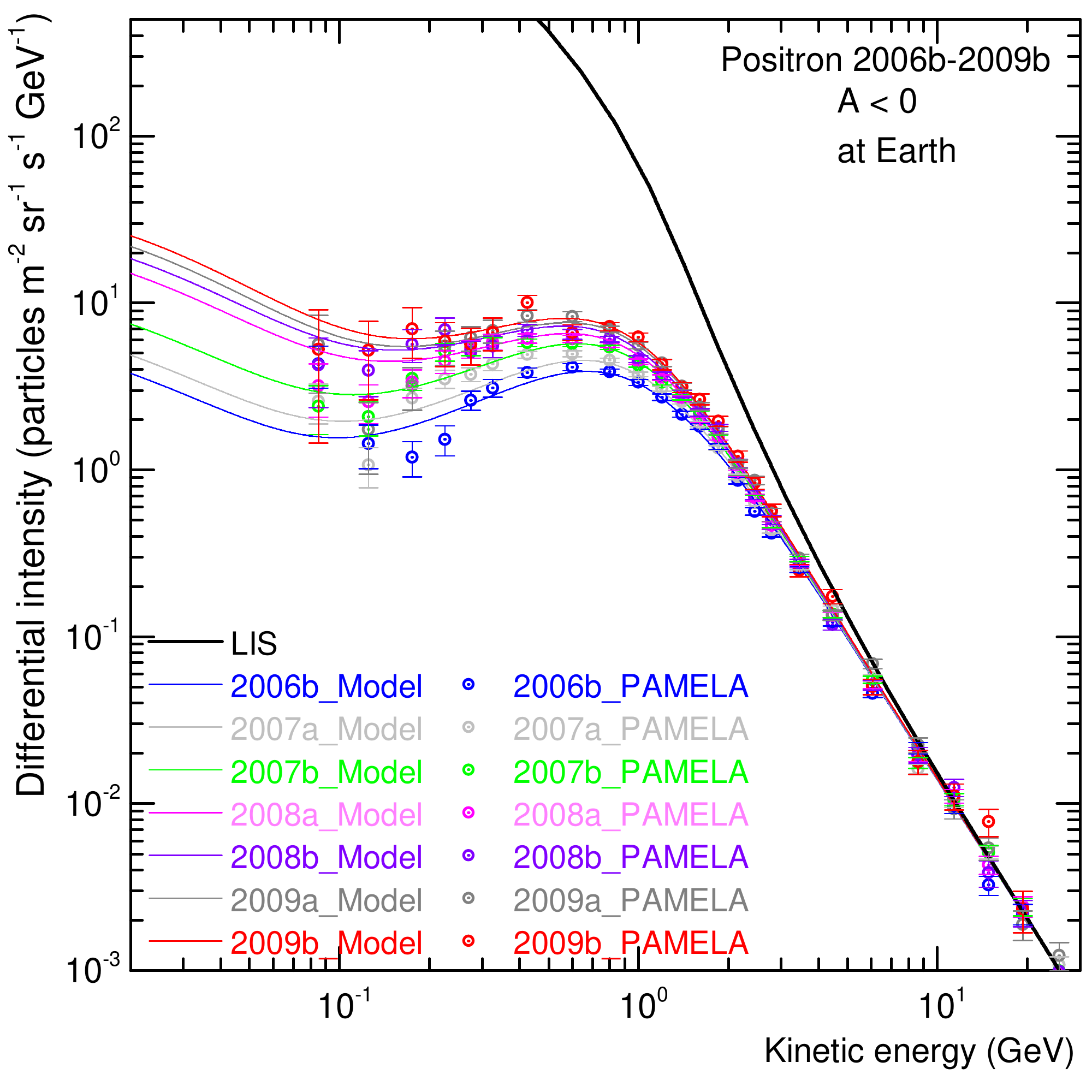}
\caption{Computed positron spectra for seven six-month averaged periods are shown from 2006b to 2009b at Earth (coloured solid lines), with respect to the LIS at 122 au (solid black line), together with the corresponding observed $\it{PAMELA}$ positron spectra (coloured points) from Figure 1. \label{fig7}} 
\end{figure}

Figure \ref {fig7} shows the observed positron spectra from Figure 1 (coloured circles) overlaid by the corresponding computed spectra (coloured solid lines) at the Earth with respect to the LIS at 122 au. The modulation of positrons becomes significant below 30 GeV, and increases gradually with decreasing energies, similar to other CR particles. The effect of continuously varying heliospheric modulation conditions on galactic positrons is also evident from below 10 GeV. The positron spectrum for July–December 2006 (2006b; blue line) is affected the most, decreasing gradually up to July-December, 2006 (2009b; red line) in accord with the gradual decrease in solar activity. These modulated positrons exhibit a characteristic peak in each spectrum just below 1 GeV in all seven semesters. The kinetic energy where the peaks occur gradually shifts to lower values with decreasing modulation as more low energy positrons reach the Earth. The spectra decrease with lowering energies to become a minimum around 100-200 MeV; for 2006b this energy value is meaningful lower than for 2009b. Below 100 MeV the spectra turn up as the modulation becomes diffusion dominated. The match between the computed and observed spectra is most reasonable down to about 200 MeV but clearly less so at lower energies. 

The computed intensity ratio of each semester with respect to 2006b (solid lines), along with the observed positron intensity ratio with respect to 2006b (data points with error bars) adopted from Figures 2 and 7 are shown in Figure 8. Consistent with the results in the previous figure, the intensity ratio with respect to 2006b stays $\approx$1 down to about 20 GeV, then to deviate quickly and significantly below $\approx$5 GeV. The confidence level of the observed positron spectra below 200 MeV is relatively low as the total error bars indicate. The intensity ratio increases gradually up  a maximum value around 30-40 MeV. Below this energy, it shows a continuous relatively small decrease. As expected, the ratio with respect to 2006b is highest for 2009b (end of the solar activity minimum) and it decreases for each semester (2009a to 2007a), to unity for 2006b. For example, at 30 MeV the intensity of 2009b is increased by a factor of $\approx$6.5 with respect to 2006b, at 100 MeV it was $\approx$4.5 and at 1 GeV it was $\approx$2.0. 
\begin{figure}[!htp]
\plotone{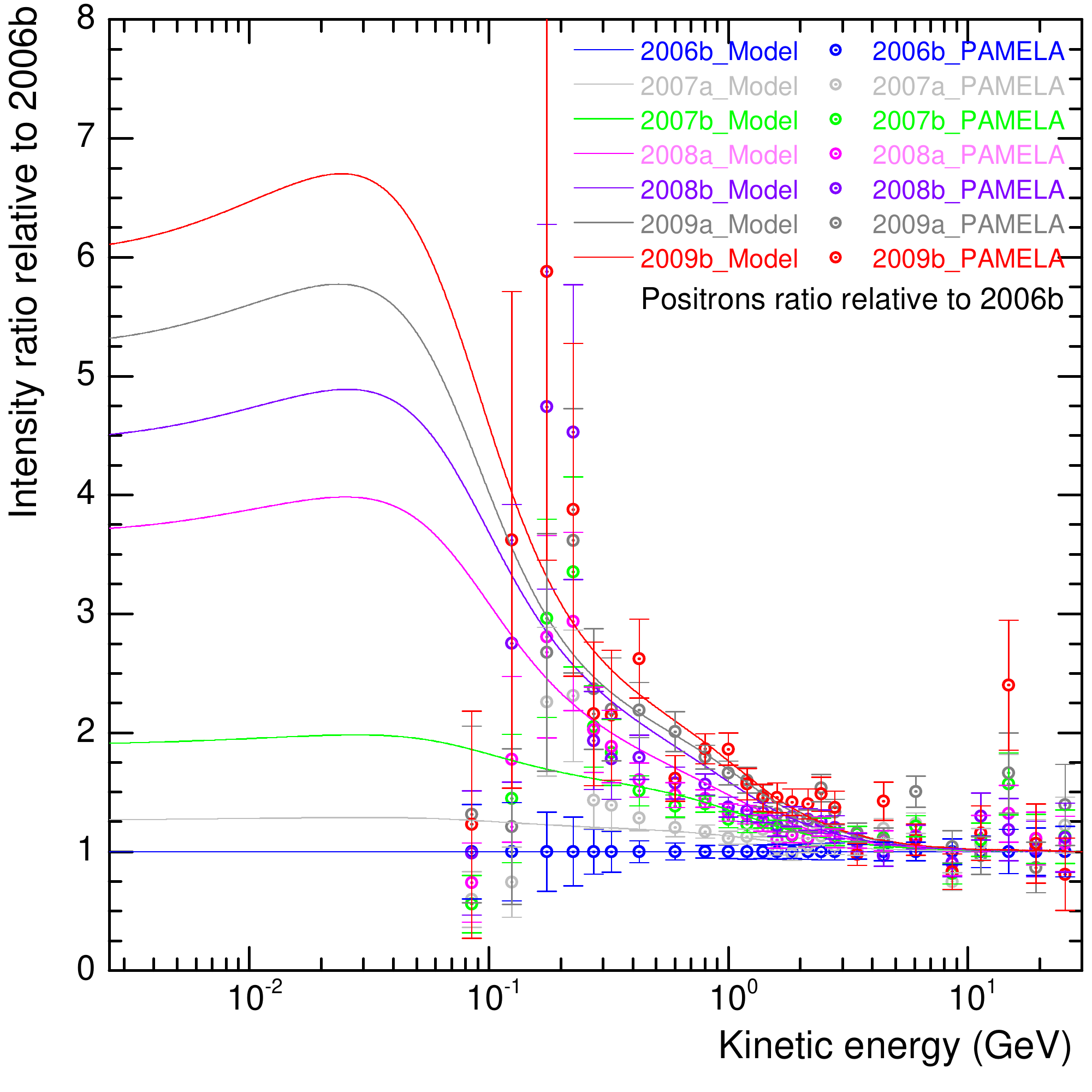}
\caption{Computed positron intensity ratio with respect to the values in 2006b for each semester (solid lines) along with the observed intensity ratios also with respect to 2006b (data points), adopted from Figures 2 and 7. As expected, the ratio is the highest for 2009b (red solid line).
\label{fig8}}
\end{figure}

The ratio between the positron LIS and the modulated spectra at Earth tell how much modulation happens inside the heliosphere; this ratio is called the modulation factor (MF) as shown in Figure 9. The MF is calculated as a function of kinetic energy for each semester by taking the ratio of the computed intensity at the Earth with respect to the LIS at 122 au. Evidently, the spectrum for 2006b is been modulated the most, with the largest MF about 0.0003 around 60 MeV. At lower energies the MF becomes steady as the modulated spectra assume the spectral shape of the LIS.
\begin{figure}[!htp]
\plotone{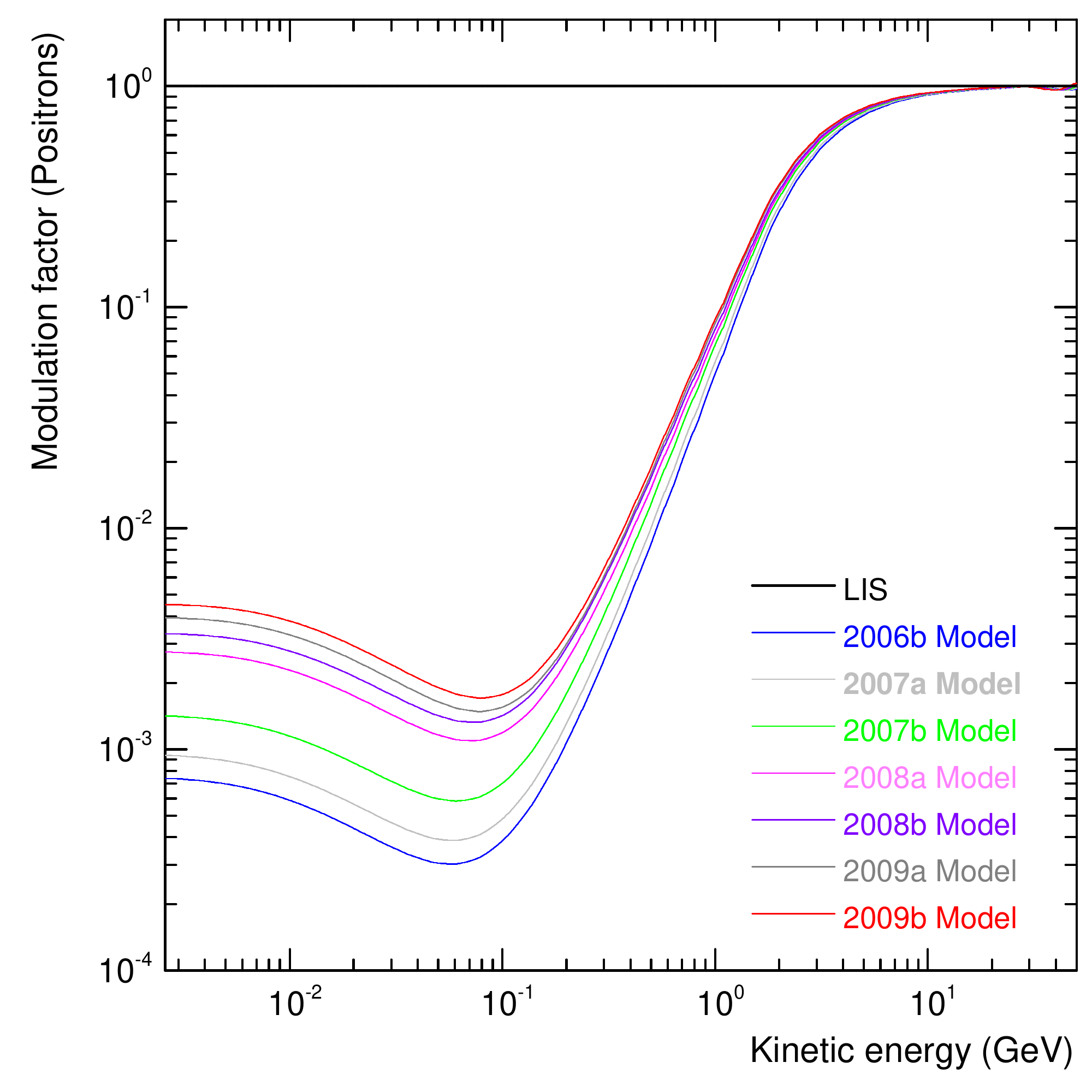}
\caption{Modulation factor at Earth calculated for positrons for each semester from 2006b to 2009b. This ratio indicates the amount of modulation occurring between the LIS value and the computed modulated spectra at the Earth. \label{fig9}}
\end{figure}

As a further illustration of positron modulation, the MF calculated for the computed modulated spectrum of 2006b is shown in Figure 10 at different radial locations, from the Earth (1 au) to 100 au from the Sun, with the heliopause at 122 au.
\begin{figure}[!htp]
\plotone{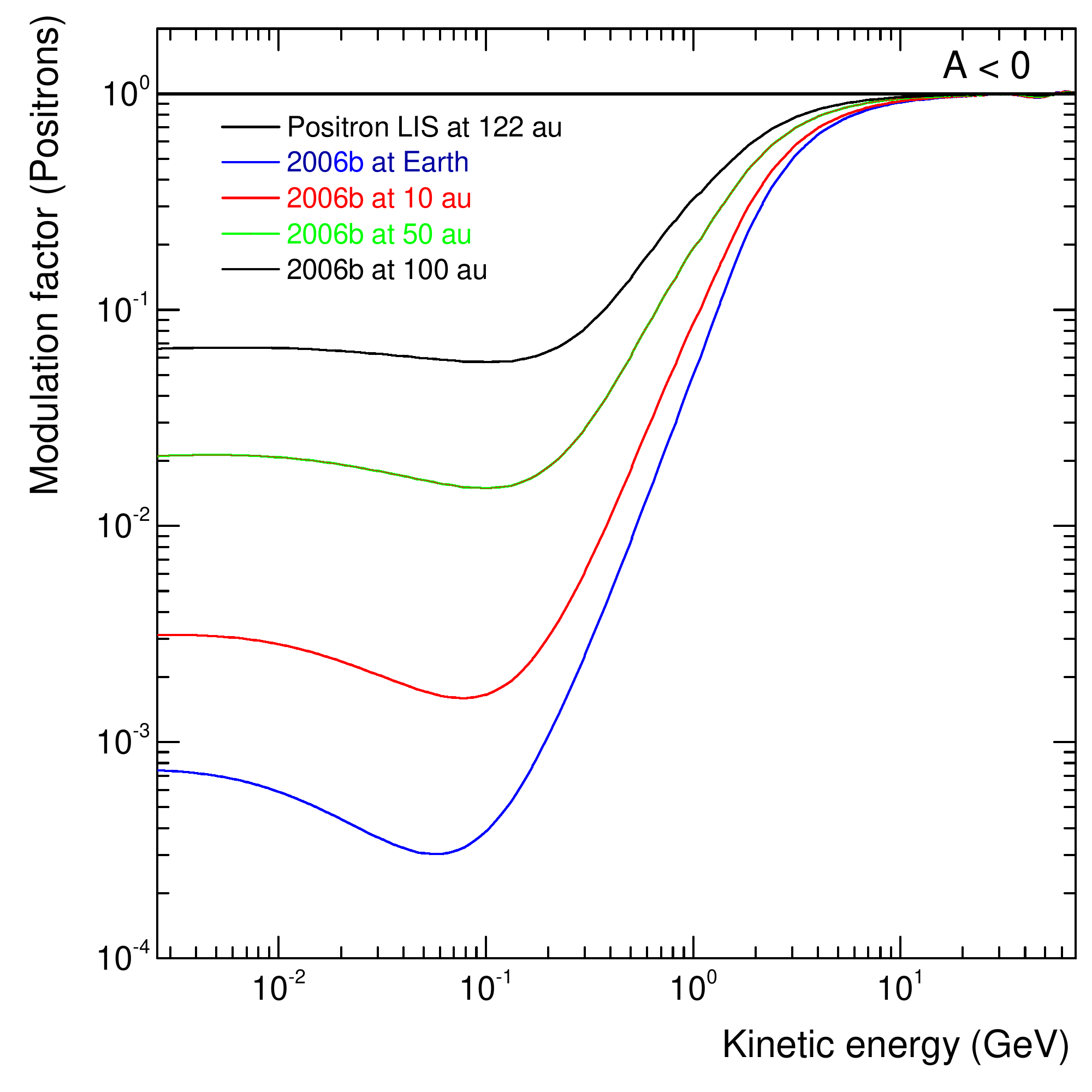}
\caption{Modulation factor calculated for the reproduced positron spectrum of 2006b, for different radial distances, at 1 au (Earth), 10 au, 50 au and 100 au.\label{fig10}}
\end{figure}
       
\subsection{The effect of drifts and HMF Polarity on Positrons} \label{sec4.3}
The solar minimum between solar cycles 23/24 is a so-called negative (A$<$0) polarity period. The Sun reversed this polarity during the solar maximum period so that it turned to an A$>$0 polarity completely in April 2014 (Sun's polar magnetic filed data source: http://wso.stanford.edu). According to drift theory, when this happens, the drift pattern of both positively and negatively charged CR particles change direction; during an A$>$0 cycle, the positrons drift inward towards the Earth mainly through the polar regions of the heliosphere and then outwards mostly along the HCS.
\begin{figure}[!htp]
\plotone{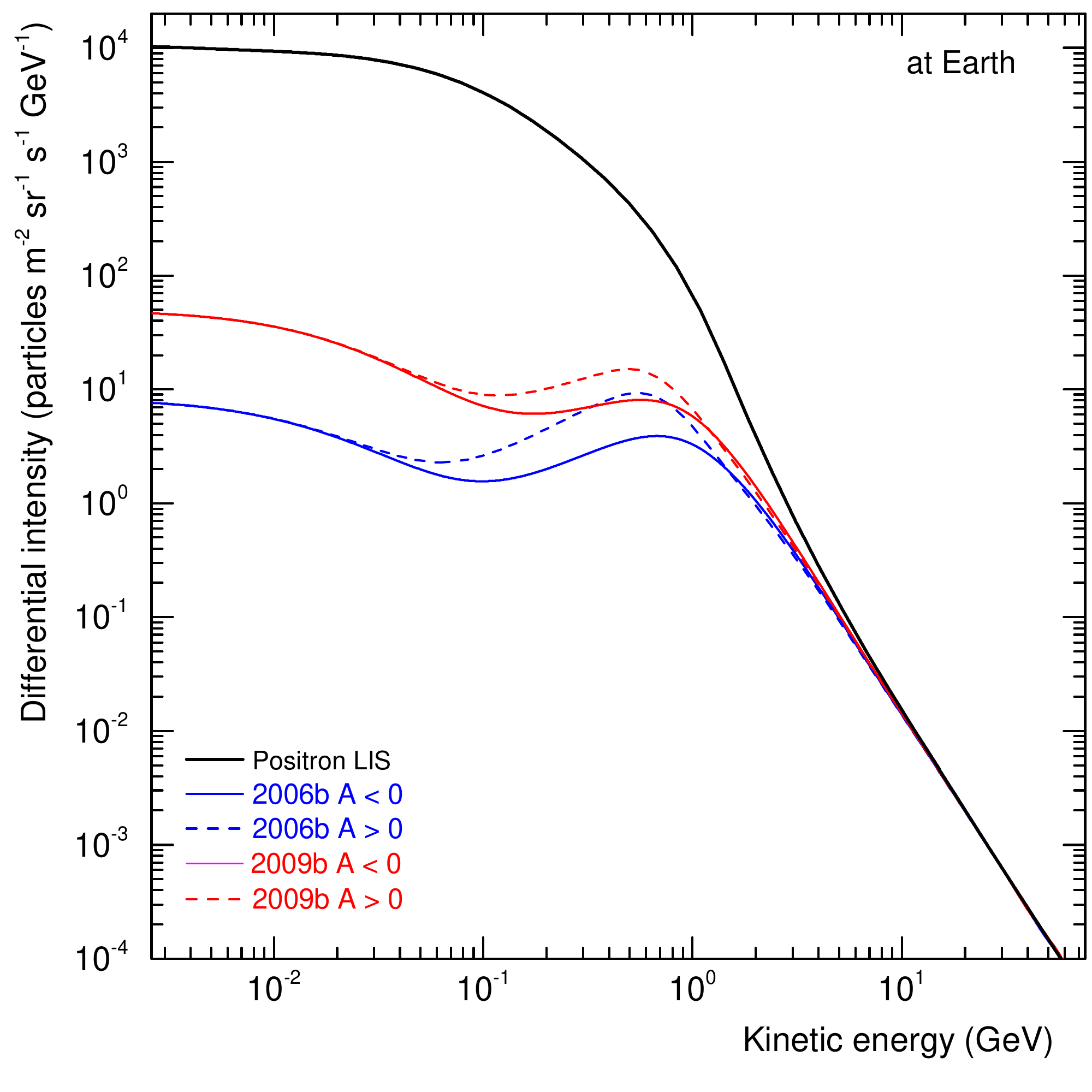}
\caption{Predicted positron spectral intensity (dashed lines) at the Earth, with respect to the LIS at 122 au, for an A$>$0 polarity epoch compared to the spectra for 2006b and 2009b of the A$<$0 epoch (solid lines), using the same  modulation conditions except for the change in the HMF direction and the subsequent change in drift directions.\label{fig11}}
\end{figure}

Figure 11 illustrates what happens to the modulated positron spectra at the Earth when this HMF direction reversal is implemented in the model. It shows the computed positron spectra for 2006b (blue solid line) and 2009b (red solid line), both being in an A$<$0 epoch, compared to the corresponding spectra for an A$>$0 epoch (blue and red dashed lines), assuming similar modulation conditions, except that only the HMF direction was changed and consequently the drift directions. Evidently, higher spectral intensity for positrons is predicted during this A$>$0 period compared to the A$<$0 period, down to about 20 MeV where the modulation process become diffusion dominated for positrons (see Figure 6) and the drift effects subside. Note that for the modulated spectra above 10 GeV, drift effects are still there but become increasingly insignificant to gradually subside. In the case of positrons, drift effects are a maximum between about 100 MeV to 2 GeV, depending on the level of modulation.       

We are interested also in looking at how the positron spectra vary radially from the Earth to the heliopause during both A$<$0 and A$>$ 0 polarity. This is illustrated in Figure 12, showing the computed positron intensity at different radial distances from 1 au to the heliopause at 122 au during the 2006b period; at 1 au (Earth), 10 au, 50 au and 100 au (coloured solid lines). The coloured dashed lines are the corresponding computed positron intensity for an A$>$ 0 polarity epoch with similar modulation conditions as in 2006b.  
\begin{figure}[!htp]
\plotone{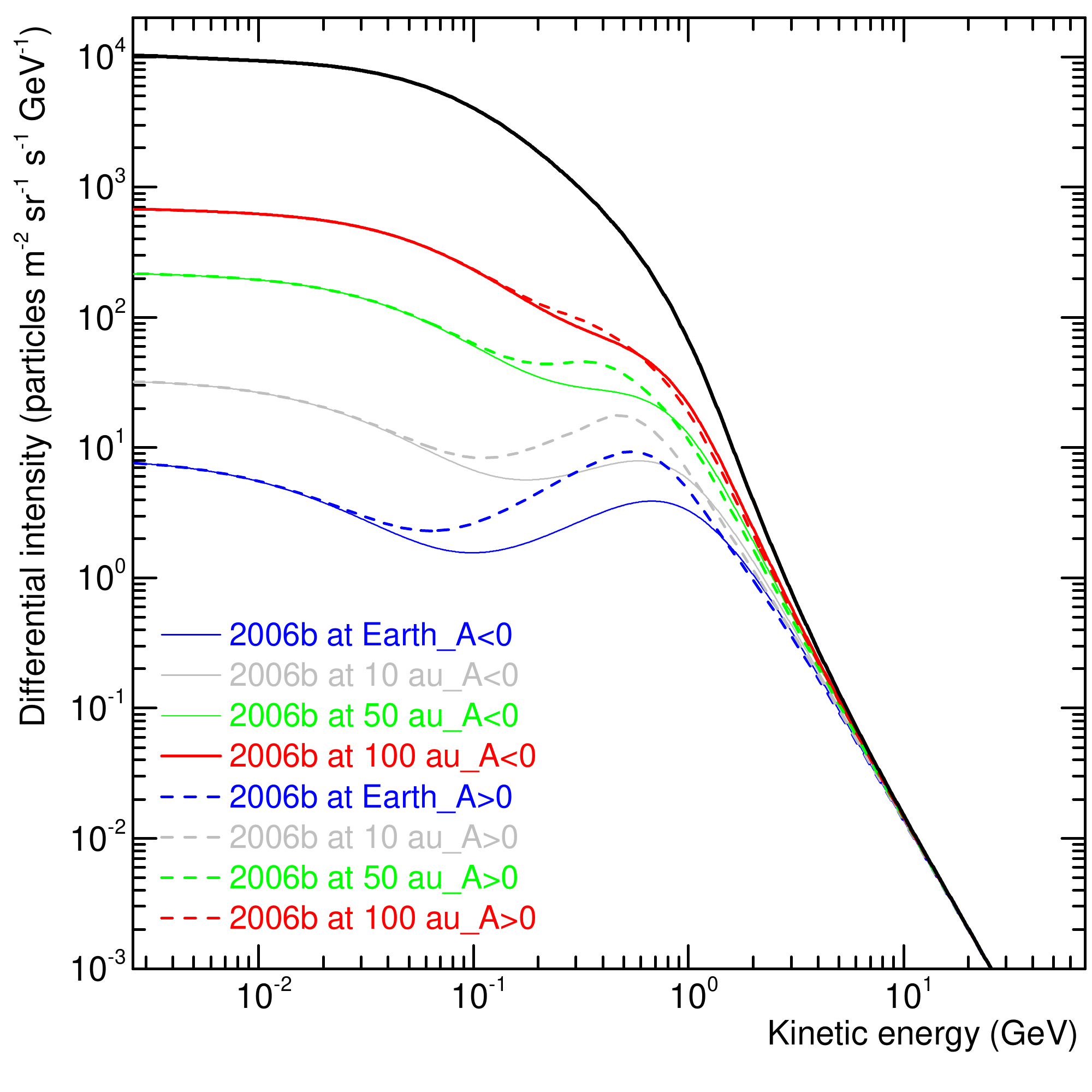}
\caption{Computed positron spectra for 2006b at different radial distance, at Earth, 10 au, 50 au and 100 au (colored solid lines), with respect to the LIS at 122 au. The colored dashed lines are the corresponding spectra for the A$>$0 epoch with similar modulation conditions as in 2006b, except for the change in drift directions.\label{fig12}}
\end{figure}

\begin{table}
\centering
\caption{Summary of the calculated intrinsic and modulation parameters used to reproduce the positron measurements at 1 GV from PAMELA for the period July 2006 to December 2009. \label{Table1}} 
\begin{tabular} {c c c c c c c c}
\hline \hline
Parameters & 2006b & 2007a & 2007b & 2008a & 2008b & 2009a & 2009b \\
\hline
$\alpha (^{\circ})$ & $16.80$ & $14.43$ & $14.08$ & $15.56$ & $14.38$ & $11.01$ & $9.50$ \\
$B_{e}$ (nT) & $4.95$ & $4.72$ & $4.36$ & $4.27$ & $4.11$ & $3.97$ & $3.91$ \\
$ r_{TS}$  (AU)  & $88.0$ & $87.0$ & $86.0$ & $85.0$ & $84.0$ & $82.0$ & $80.0$ \\
\hline
$\lambda_{\parallel}$ (AU) & $0.438$ & $0.465$ & $0.506$ & $0.539$ & $0.557$ & $0.574$ & $0.593$ \\
$K_{A0}$  & $0.90$ & $0.90$ & $0.90$ & $0.90$ & $0.90$ & $0.90$ & $0.90$ \\
$P_{A0}  (GV)$  & $0.90$ & $0.90$ & $0.90$ & $0.90$ & $0.90$ & $0.90$ & $0.90$ \\
$K_{\perp r}^{0}$  & $0.02$ & $0.02$ & $0.02$ & $0.02$ & $0.02$ & $0.02$ & $0.02$ \\
$K_{\perp \theta}^{0}$  & $0.02$ & $0.02$ & $0.02$ & $0.02$ & $0.02$ & $0.02$ & $0.02$ \\
$c_{1}$  & $0.00$ & $0.00$ & $0.00$ & $0.00$ & $0.00$ & $0.00$ & $0.00$ \\
$c_{2 \parallel}$  & $2.25$ & $2.25$ & $2.25$ & $2.25$ & $2.25$ & $2.25$ & $2.25$ \\
$c_{2 \perp}$  & $1.688$ & $1.688$ & $1.688$ & $1.688$ & $1.688$ & $1.688$ & $1.688$ \\
$c_{3}$  & $2.50$ & $2.50$ & $2.50$ & $2.50$ & $2.50$ & $2.70$ & $2.70$ \\
$P_{k}  (GV)$  & $0.585$ & $0.575$ & $0.565$ & $0.58$ & $0.58$ & $0.57$ & $0.57$ \\
$d_{\perp \theta}$  & $6.00$ & $6.00$ & $6.00$ & $6.00$ & $6.00$ & $6.00$ & $6.00$ \\
\hline
\end{tabular}
\end{table}

The drift effect on positrons decreases gradually from Earth to the heliopause. The difference between the A$>$0 and A$<$0 spectra is a maximum at Earth, decreasing with increasing radial distance, becoming much less at 100 au. In terms of energy, the maximum difference shifts to higher energies with increasing radial distances. Note that the computed A$>$0 and A$<$0 spectra cross between 1 and 2 GeV at the Earth. This cross-over energy shifts to lower energies with increasing radial distance. This effect is required to explain why at NM energies the intensity during A$>$0 epochs is lower than during A$<$0 spectra in contrast to what occurs at lower energies; for a detailed discussion of this intriguing effect for Galactic protons, see Reinecke \& Potgieter (1994). As seen before, drift effects vanish at lower energies whereas at high energies the effects gradually subside.

\section{Discussion} \label{sec5}
The primary objective of this study is the reproduction of six-month averaged positron spectra as observed by $\it{PAMELA}$ from July 2006 to December 2009. This is accomplished by using a well-established and comprehensive 3D numerical modulation model, which had been used previously to study the heliospheric modulation of monthly averaged Galactic proton spectra observed by $\it{PAMELA}$ for the same period, over the energy range 80 MeV to 30 GeV (Potgieter et al. 2014; Vos \& Potgieter 2015; Potgieter \& Vos, 2017), and for the modulation of six-month averaged Galactic electron spectra observed by $\it{PAMELA}$ for the same period, over the same energy range which is important for solar modulation (Potgieter et al. 2015; Potgieter \& Vos, 2017).

When simulating the modulation of CRs, the tilt angle ($\alpha$) of the HCS and HMF strength at Earth ($B_{e}$) are considered very good proxies for solar activity. The moving average values of $\alpha$ and $B_{e}$ are used as the essential time-varying modulation parameters, used as input to the numerical model along with the position of the termination shock to reproduce the positron $\it{PAMELA}$ spectra; these parameters are shown in the first three rows of Table 1. 
Changing $B_{e}$ affects directly the value of the diffusion and drift coefficients as in Equations (9), (14) and (15). However, in order to reproduce the observed $\it{PAMELA}$ positron spectra, the value of the diffusion coefficients had to be increased for each semester from 2006b to 2009b, that is, in addition to the changes in the three mentioned parameters. A similar result was reported by Potgieter et al. (2014) for the proton spectra.

The ratio of the perpendicular diffusion coefficients to the parallel diffusion coefficient is kept at 0.02, which gives $K_{\perp r}$ = $K_{\perp \theta} = $0.02 $K_{\parallel}$, but an additional enhancement of a factor of $d_{\perp \theta}$ = 6 for $K_{\perp \theta}$ in the polar direction is also required. We increased the value of $\lambda_{\parallel}$, and consequently also $\lambda_{\perp r}$ and $\lambda_{\perp \theta}$ by a factor of $\approx$1.35 from 2006b to 2009b, as shown in the fourth row of Table 1.

The value of $K_{A0}$ = 0.90 and $P_{A0}$ = 0.90 GV for all seven periods means we reduced drift in an implicit way through increasing $K_{\parallel}$, $K_{\perp r}$ and $K_{\perp \theta}$, without changing the $K_{D}$ (in Equation 9). Furthermore, $c_{1}$ together with $c_{2}$  ($c_{2 \parallel}$ and $c_{2 \perp}$) determines the slope of rigidity dependence above and below of $P_{k}$. 
 
In the case of positrons, $c_{1}$ = 0, $c_{2 \parallel}$ = 2.25, and $c_{2 \perp}$ = 1.688 (0.75$c_{2 \parallel}$) are kept constant for all seven periods; only the value of $P_{k}$ is changed when required, to reproduce the seven positron spectra. The smoothness of the transition in rigidity dependence is determined by $c_{3} = 2.50 $ for 2006b to 2008b and  $c_{3} = 2.70 $ for 2009a and 2009b as shown in Figure 6 and listed in Table 1.                

Although particle drift played a significant role (Di Felice et al. 2017; Munini et al. 2018) in the modulation of electrons and positrons during the unusual 2006-2009 minimum period, the process was diffusion dominated. As such, an unexpected result for an A$<$0 solar minimum epoch (phase). Drift effects below 20 MeV and above 10 GeV are found to be negligible. The drift effect at Earth is found to be the lowest during the July- December 2009 period (when solar activity was also at its lowest). 

Figures 10 \& 12 illustrate the total modulation of positrons from the heliopause to the Earth. Our 3D model predicts that the intensity of positrons, over the energy range 1 MeV to 10 GeV, will be higher during the upcoming A$>$0 solar minimum period than during 2006-2009, assuming similar quiet modulation conditions.           

The next phase of this study is to apply this model to Galactic electron and positron spectra observed by $\it{AMS}$-02 (Aguilar et al. 2018) for the period 2011-2017, as was done for $\it{PAMELA}$ protons (Martucci et al. 2018). These $\it{AMS}$-02 observations together with the $\it{PAMELA}$ observations for 2006-2009, can provide electron and positron spectra over a full solar cycle.  It will help to understand how the modulation of positrons is different from that for electrons over a complete solar cycle. This may also clarify the role of drift over a complete solar cycle, especially how large drift affects are during a HMF reversal period (see Adriani et al. 2016).          

\section{Summary and Conclusions} \label{sec6}

The six-month averaged cosmic ray positron spectra of energy range 80 MeV to 30 GeV, recorded by $\it{PAMELA}$ from July 2006 to December 2009 (Munini 2015) were used together with a comprehensive three-dimensional numerical modulation model, which includes particle drift and a solar cycle dependent dynamic heliosheath, to understand how the solar modulation of galactic positrons is affected by the unusual solar minimum conditions of solar cycles 23/24. For this purpose, a new very LIS for Galactic positrons was constructed. 

It follows from the observations that the positron intensity has increased gradually from July-December 2006 to have reached its highest level during July-December 2009 in accord with decreasing solar activity. This recovery towards solar minimum modulation is effectively reproduced by the model using the same approach as for the modulation of protons and electrons observed by $\it{PAMELA}$. An effort was made to keep the modeling approach as simple as possible within the context of a 3D drift model but to adhere to the main features of CR modulation as observed, also by previous space missions such as Ulysses. Essentially all three the diffusion coefficients scale as 1/B, with a small modification over the first 5 AU for the radial dependence of $K_{\parallel}$ but a much larger modification to $K_{\perp \theta}$ in terms of its polar dependence. Apart from changing the tilt angle of the HCS, the HMF magnitude and the position of the termination shock in a time-varying manner, we found that it was necessary to increase additionally e.g. $\lambda_{\parallel}$ at the Earth from 0.438 au to 0.593 au at 1 GV in order to reproduce the positron spectra from 2006b to 2009b. 

Comparison of simulated spectra with precise observations makes it possible to extend the modeling with confidence down to 1 MeV for positrons where observations are unavailable. It follows that the LIS becomes flat, essentially independent of energy below 50 MeV. Because positron modulation is dominated by diffusion at these low energies (see Figure 6), similar to electron modulation, the spectral shape of the LIS at these energies is maintained to deep inside the heliosphere (see Figures 4, 11, \& 12). This implies that if the positron spectrum could be observed at these energies at the Earth, it would tell what the spectral shape of the very LIS is. This is in sharp contrast to protons, anti-protons and all CR nuclei, the spectra of which are determined by adiabatic energy losses at low energies at the Earth.
 
It is interesting to note that the highest positron spectrum was observed by $\it{PAMELA}$ during the second half of 2009 (July-December 2009), similar to Galactic electrons and protons (highest intensity in December 2009), for the latter the highest since the beginning of the space era. This was unexpected for positively charged CR particles (protons, positrons, etc.) during an A$<$0 polarity epoch because drift theory generally predicts lower spectra for positively charged CRs during A$<$0 cycles. As such the unusual 2006-2009 minimum solar activity period seems to have broken all predictions and expectations caused by the Sun that unexpectedly became relatively very quiet and seems to continue to be so. If similar modulation conditions would prevail during the upcoming solar minimum period, an A$>$0 cycle with different drift patterns than during 2006-2009, the spectrum for protons, positrons and all heavy nuclei should even be higher than in December 2009. If the Sun is even quieter than in 2006-2009, new record CR intensities will then be set.
             
\acknowledgments The authors from SA express their gratitude for the partial funding by the South African National Research Foundation (NRF) under grant 98947. The authors wish to thank the GALPROP developers and their funding bodies for access to and use of the GALPROPWebRun service. We also acknowledge the use of HCS tilt data from Wilcox solar observatory's http://wso.stanford.edu webiste, and HMF data from NASA's OMNIWEB data interface http://omniweb.gsfc.nasa.gov. OPMA \& DB acknowledges the partial financial support from the post-doctoral program of the North-West University, South Africa. OPMA also thanks Jan-Louis Raath for many useful discussions.

\end{document}